\documentclass[%
 reprint,
 amsmath,amssymb,
 aps,
]{revtex4-1}

\usepackage{subcaption}
\usepackage{graphicx}
\usepackage{dcolumn}
\usepackage{bm}

\begin{document}
\captionsetup[figure]{justification=centerlast, font=footnotesize}

\preprint{APS/123-QED}

\title{Optimal mode configuration for multiple phase-matched four-wave mixing processes}

\author{Erin M. Knutson}
\author{Jon D. Swaim}
\author{Sara Wyllie}
\author{Ryan T. Glasser}
 \email{Corresponding author: rglasser@tulane.edu}
\affiliation{Department of Physics, Tulane University, New Orleans, LA USA 70118}

\date{\today}

\begin{abstract}
We demonstrate an unseeded, multimode four-wave mixing process in hot $^{85}$Rb vapor, using two pump beams of the same frequency that cross at a small angle. This results in the simultaneous fulfillment of multiple phase-matching conditions that reinforce one another to produce four intensity-stabilized bright output modes at two different frequencies. Each generated photon is directly correlated to exactly two others, resulting in the preferred four-mode output, in contrast to other multimode four-wave mixing experiments. This provides significant insight into the optimal configuration of the output multimode squeezed and entangled states generated in such four-wave mixing systems.  
We examine the power, temperature and frequency dependence of this new output and compare to the conical four-wave mixing emission from a single pump beam.  The generated beams are spatially separated, allowing a natural distribution for potential use in quantum communication and secret-sharing protocols.
\end{abstract}

\maketitle

\section{\label{sec:level1}Introduction}

Multimode quantum resources are invaluable in fundamental tests of quantum physics \cite{treps_surpassing_2002,de_valcarcel_multimode_2006,chalopin_multimode_2010}, and hold many promising applications in quantum imaging \cite{treps_quantum_2003,brida_experimental_2011}, quantum metrology \cite{knott_local_2016,humphreys_quantum_2013} and quantum communication \cite{chalopin_direct_2011,lassen_tools_2007,wen_triple-mode_2016}. Much recent work involves new four-wave mixing (FWM) phase-matching geometries  in order to generate many quantum-mechanically correlated beams \cite{slusher_observation_1985,zhang_dressed_2015,boyer_entangled_2008,pooser_quantum_2009}, either using multiple pump beams \cite{jia_generation_2017, wang_single-step_2017} or by cascading FWM setups \cite{wang_generation_2016, wang_phase-sensitive_2017, wang_characterization_2017, qin_experimental_2014, qin_experimental_2015, cao_experimental_2017, cai_quantum-network_2015, pooser_continuous-variable_2014}. In the cascaded case, the addition of one quantum-correlated mode necessitates the addition of a new vapor cell, which entails further alignment and appended loss. Additionally, all of these geometries include a probe, or seed, field shifted in frequency from that of the pump(s). 

\begin{figure}[b]
    \centering
        \includegraphics[width=\linewidth]{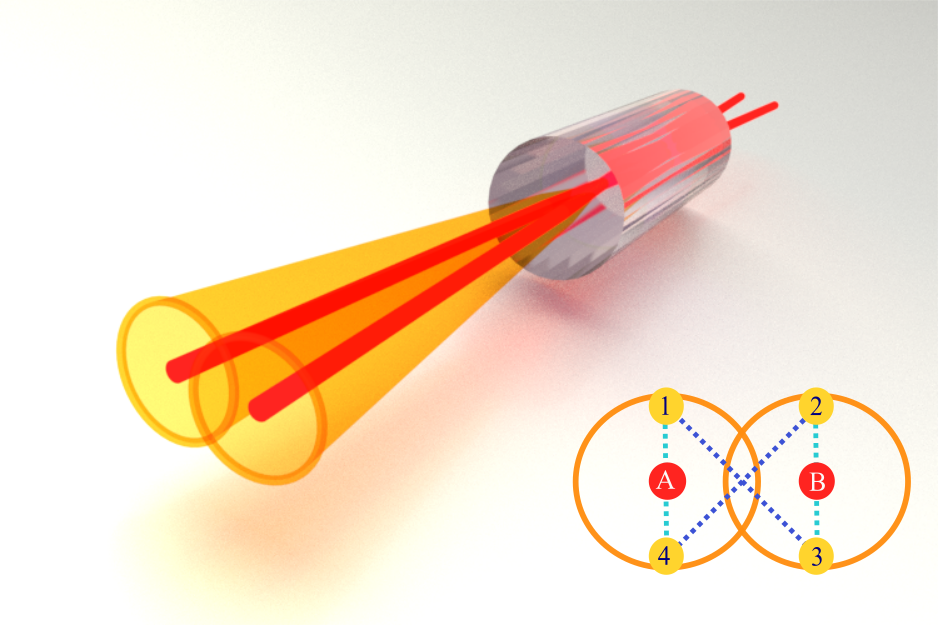}
    \caption{\textbf{Diagram of the rubidium cell with pump beams crossing at an exaggerated angle.} While each individual pump generates a four-wave mixing cone, the combination of both pumps creates four localized modes of four-wave mixing (not shown). \textbf{Inset:} Diagram showing interactions present in this configuration. The four generated modes are labeled $1,2,3,4$ clockwise starting from top left throughout this manuscript. The light blue dotted line is a single-pump FWM interaction, while the dark blue dotted line is a dual-pump FWM interaction. The orange rings are shown for scale, but are not visible simultaneously with the four modes. }
    \label{fig:setup}
\end{figure}

Here we demonstrate a self-reinforced four-mode optical stability where frequency conversion is obtained in spatially-separated modes by injecting only one laser frequency. The well-defined output modes result from the simultaneous fulfillment of multiple phase-matching conditions \cite{turnbull_role_2013, grynberg_mirrorless_1988, petrossian_transverse-pattern_1992,daems_spatial_2010} that directly reinforce one another.  This gives rise to a stable four-mode output, which is preferred over other phase-matched multimode configurations, as every FWM process directly stimulates photons in each of the output modes.  These results suggest that the optimally squeezed and entangled modes in such multi-pump-beam FWM experiments are those whose different phase-matched processes directly stimulate one another.

\begin{figure}[htbp]
\includegraphics*[width=0.7\linewidth]{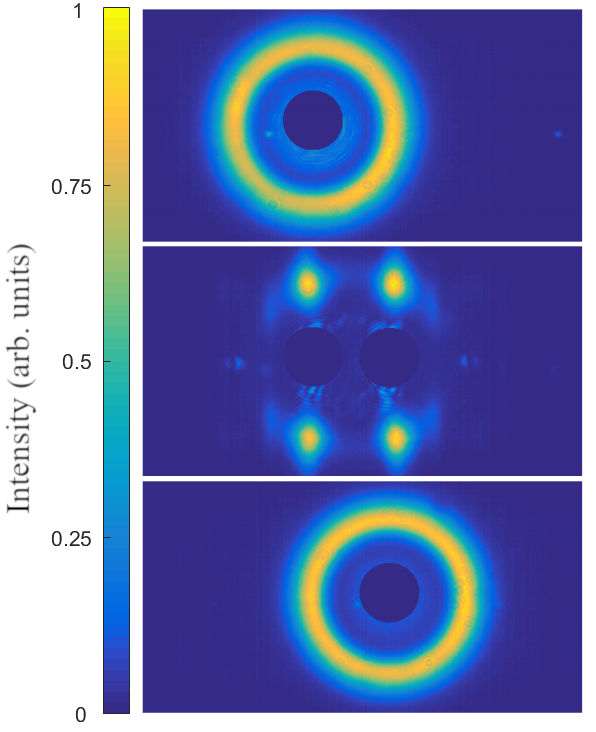}
\caption{\textbf{Camera images 150 mm after the cell of the single- and double-pump configurations, with pump areas subtracted. }\textbf{Top}: pump A only; \textbf{Middle}: both pumps; \textbf{Bottom}: pump B only. In the case of either single pump, the resultant output mode is a spontaneous FWM ring, whereas the two-pump case results in a collapse to the distinct four-mode stability.}
\label{fig:rings}
\end{figure}

\begin{figure}[htbp]
    \centering
        \includegraphics[width=\linewidth]{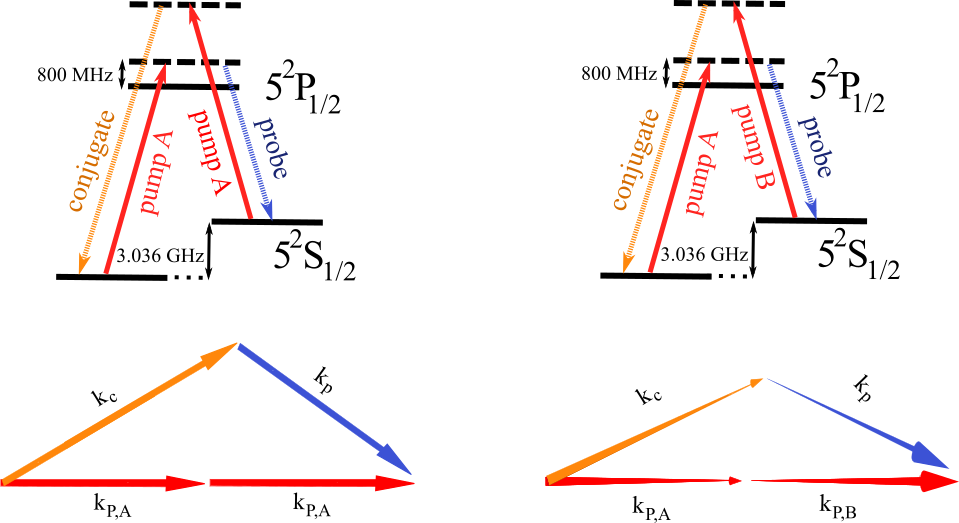}
    \caption{\textbf{Level schemes and phase matching diagrams for both processes involved in the four-mode FWM configuration. Top:} Energy level schemes for FWM on $^{85}$Rb D1 line, for the single- (left) and dual- (right) pump processes. \textbf{Bottom:} Corresponding k-vector diagrams displaying the phase-matching in each FWM process. Changing arrow thickness in the dual-pump case is used to illustrate the vectors entering and exiting the plane of the page.}
    \label{fig:k}
\end{figure}

We make use of macroscopic spontaneous FWM \cite{podhora_nonclassical_2017,gupta_multi-channel_2016,zhou_characterizing_2016,zhou_optical_2012}
by heating the nonlinear medium (rubidium vapor) to a higher temperature than in typical seeded FWM experiments. This results in the formation of a bright four-wave mixing cone about each individual pump beam, i.e. a process seeded by vacuum where opposite angular positions about each ring-shaped cross-section  of the cone are phase-matched \cite{zhou_characterizing_2016}, as shown schematically in Figure \ref{fig:setup}. An energy-level diagram as well as a phase-matching diagram for this single-pump FWM process are also shown on the left in Figure \ref{fig:k}. The two pump beams are then aligned such that they cross at a small angle (typically $\sim$0.9 degrees, the angle that optimizes output intensity), and the output is detected 150 mm after the cell. Instead of two superimposed rings centered around the pump beams, we find that the output ``collapses'' into four spatial modes, above and below each pump, as shown in Figure \ref{fig:rings}. This configuration is induced by a two-pump forward phase-matching geometry (see the right side of Figure \ref{fig:k}) that is satisfied in addition to each single-pump phase-matching condition.  Interestingly, the spontaneously generated modes form above and below each pump, rather than, e.g., at the two ``intersection points'' between the single-pump rings. The latter scenario would ostensibly result in a six-mode formation as in \cite{wang_single-step_2017}, in which an input probe was seeded at one of these intersection points. Instead, however, the four-mode output arises because the dual-pump phase-matching geometry directly stimulates two single-pump phase-matched FWM processes (and vice versa), resulting in each photon being directly correlated to exactly two others, as illustrated in the inset in Fig. \ref{fig:setup}.  Thus, each of the four FWM processes (two single-pump and two dual-pump processes) reinforce one another, and the four-mode output is realized.  When the pump powers are balanced, the output mode profile is highly stable and symmetric. This new configuration boasts a large amount of tunability in the relative power of each pump, total pump power, atomic vapor temperature, pump frequency, and angle between pumps. It may be applied, for example, to multi-channel entanglement distribution using spatial multiplexing as in \cite{gupta_multi-channel_2016}, where researchers generate independent pairs of spatially-separated, correlated random bit streams to be used as secure keys in a many-party secret sharing scheme.

\section{\label{sec:Exp}Experimental layout}
\begin{figure}[tp]
\begin{subfigure}{0.4\textwidth}
       \includegraphics[width=\linewidth]{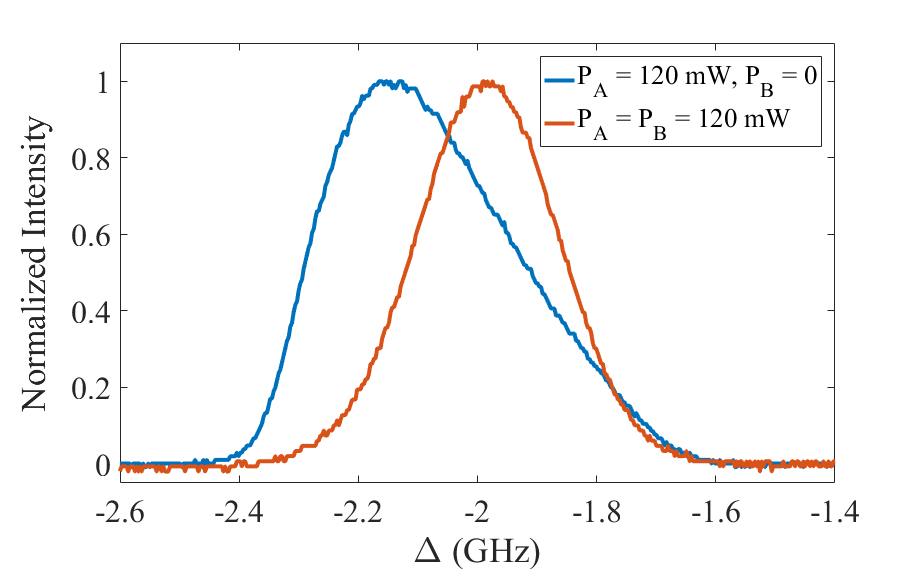}
    \end{subfigure}
 \\
    \begin{subfigure}{0.4\textwidth}
      \includegraphics[width=\linewidth]{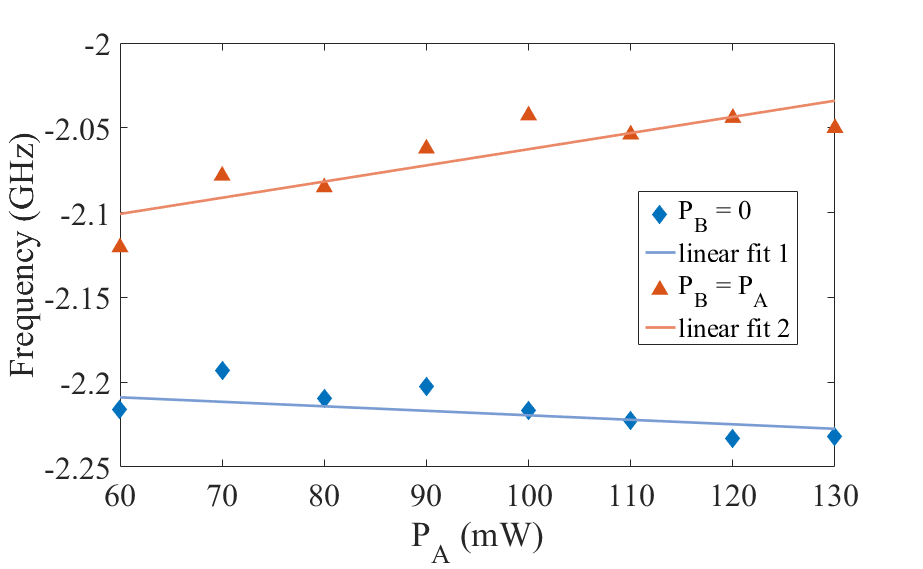}
    \end{subfigure}
\caption{\textbf{Frequency shift of intensity curves for the ring and four modes (single and dual pumps, respectively).} The addition of a second pump changes the optimal pump detuning $\Delta$, and this shift increases linearly with power. \textbf{Top:} Normalized total intensity for the ring around one pump (blue) and the four modes around two pumps (orange), as a function of overall pump detuning $\Delta$. \textbf{Bottom:} The frequency at which maximum intensity occurs for the ring (blue diamonds) and for the four modes (orange triangles), as a function of the power of pump A.}
\label{fig:pumpdetuning}
\end{figure}

Coherent light from a CW titanium-sapphire laser is tuned near the D1 line of rubidium and coupled into a single-mode, polarization-maintaining optical fiber to maintain a clean Gaussian profile ($1/e^2 \sim 900 \mu $m), then split on a polarizing beam splitter (PBS). A half-waveplate is placed in one of the beam paths in order to ensure that the pumps have the same linear polarization. The beams are then directed into the rubidium cell with a small separation angle, typically just under 1 degree. The 25.4 mm-long cell is maintained at a temperature of 145 degrees Celsius, hot enough such that a bright spontaneously-seeded ring forms about each single pump, individually. Another PBS is placed directly after the cell to filter the leftover pump light. At a distance of 150\,mm after the cell, the resulting output modes are imaged with a CCD camera (2048 x 1088 pixels with a 5.5\,$\mu$m pixel pitch). An experimental schematic is shown in Fig. \ref{fig:setup} (a). After imaging the output onto a CCD, the saturated pump areas are subtracted from each image, and the resultant pixel values are integrated over to calculate the total intensity in the areas of interest (either the four modes, or the leftover rings, or both). Additionally, for the detuning measurements shown in Figure \ref{fig:pumpdetuning} leftover scattered pump light is filtered out with an iris, selecting out the modes of interest while scanning the Ti:Sapph laser frequency.

\section{\label{sec:res}Results}
\begin{figure}[htbp]
\centering
\includegraphics*[width=\linewidth]{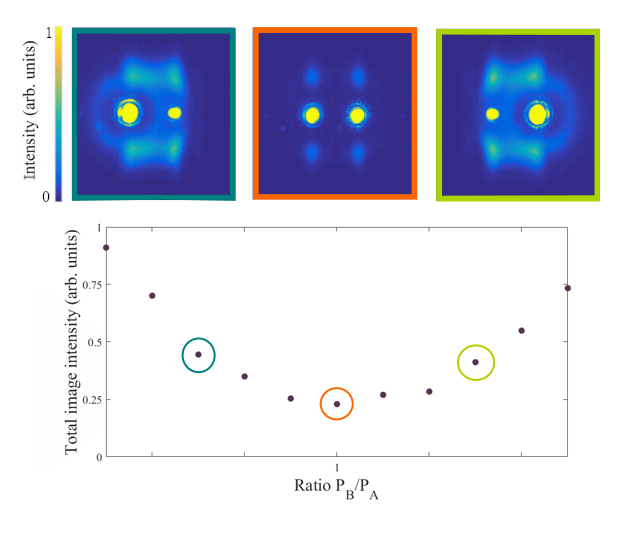}
\caption{\textbf{Top}: Camera images as the power ratio of $P_A$ to $P_B$ is varied. Blue: $P_A=143$ mW, $P_B=23$ mW; Orange: $P_A=78$ mW, $P_B=86$ mW; Green: $P_A=23$ mW, $P_B=149$ mW. Here the pumps are not subtracted from the images for visualization purposes, but when calculating intensity the pump areas are always subtracted from each image, as they saturate the camera. \textbf{Bottom}: Total intensity of generated light versus ratio of powers $P_A$ to $P_B$. The pumps are subtracted from each camera image, i.e. this measurement only takes generated light into account. \vspace{10mm}} 
\label{fig:ratiograph}
\end{figure}

\begin{figure}
\centering
\includegraphics[width=\linewidth]{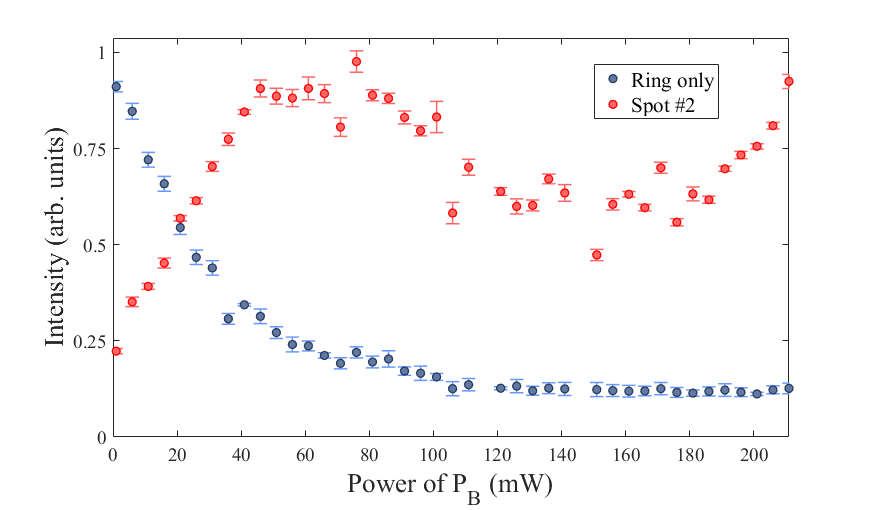}
\caption{\textbf{Total intensity of the generated light as $P_B$ is varied, with different areas excluded.} The pumps are subtracted from every image, and $P_A$ is fixed at 210 mW. The two curves are normalized individually, and the error bars represent one standard deviation and are based off of ten images taken for each $P_B$. \textbf{Red}: Intensity of generated light in mode 2 only. Interestingly, there is a local intensity maximum before $P_B=P_A=210$ mW, at $P_B\approx0.3 P_A$. \textbf{Blue}: Intensity of generated light in the area excluding the four modes, i.e. ``ring'' light only.} 
\label{fig:ringspotintens}
\end{figure}

\begin{figure}
\centering
\includegraphics*[width=\linewidth]{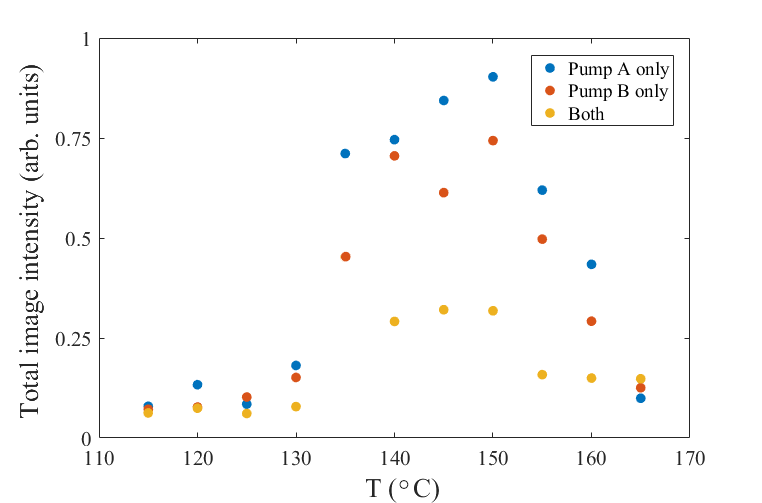}
\caption{\textbf{Effects of changing cell temperature on the total intensity of the generated modes, for pump A only (blue), pump B only (red) and both pumps on (yellow).} Pump powers $P_A$ and $P_B$ are equal, i.e. when imaging with both pumps on, the total pump power in the cell is $2P_A$.}
\label{fig:temp}
\end{figure}

We find that the four-wave mixing cones generated from each single pump ``collapse'' into an optical stability of four spatially-distinct modes when both pumps are present. We also demonstrate a shift in the intensity curve of the four modes compared to the ring as a function of overall pump detuning, as shown in Figure \ref{fig:pumpdetuning}\,(a). Furthermore, this shift increases linearly as a function of pump power, as shown in Figure \ref{fig:pumpdetuning} (b). 
The total integrated intensity of the four modes is invariably less than the two rings, indicating that the processes involving both pumps are less efficient than each single-FWM process. Figure \ref{fig:ratiograph} shows the parabolic decrease in intensity between the single- and dual-pump FWM processes, as the ratio of pump powers is changed. Note that there a minimum in total output mode intensity when the pump powers are nearly equal.
Additionally, the intensity dependence on the power of pump B, or $P_B$, while $P_A$ remains fixed, is shown in Figure \ref{fig:ringspotintens} for a single isolated mode, as well as the residual ring (the area excluding the four modes). Notably, there is a local intensity maximum in the former at approximately $P_B=\frac{1}{3}P_A$, in addition to the expected maximum at $P_B=P_A$. It is worth noting that it is difficult to fully isolate either the single- or dual-pump FWM output in this crossed-pump geometry, as they are either both present or both absent at any given cell temperature, laser power, and laser frequency. However, the first local maximum apparent in Figure \ref{fig:ringspotintens} is suggestive of a separability between the single- and dual-pump processes, since this deviates from the single-pump case wherein increased laser power results in a steady quadratic increase in intensity \cite{wang_generation_2001}. Finally, Figure \ref{fig:temp} shows the effect of changing temperature on the total intensity of the generated light, for the case of each individual pump, as well as both pumps. An optimum occurs at roughly the same temperature for each individual pump as well as both pumps: approximately $145^{\circ}$C. As expected, at even higher temperatures (atomic densities), Doppler broadening results in stronger absorption of the generated light at the blue-detuned frequencies, resulting in a net decrease in integrated intensity.

\section{Discussion}

In the case of spontaneous single-pump four-wave mixing, phase-matching occurs at opposing points about the cross-sectional ring of emitted light. This is essentially a FWM process where the input probe (seed) and conjugate modes are vacua. The optimal angle to fulfill phase-matching in $^{85}$Rb occurs at $\approx$\,8 mrad between the pump and any point on the generated cone \cite{gupta_multi-channel_2016}. When a second pump is added at a small horizontal angle, this single FWM process competes with a new phase-matched process: one photon from each pump is annihilated, one photon is generated in mode 1, and one photon is generated in mode 3 (likewise for modes 2 and 4, where the mode numbers are defined in Figure \ref{fig:setup}). 
As in the single-pump case, these generated modes form where the phase-matching condition is met, i.e. where the total wavevector of the annihilated pump photons equals the total wavevector of the generated pair of photons. Interestingly, and perhaps counterintuitively, stabilized modes do not form at the two overlap points of the single-pump FWM cones. One might intuitively expect a total of six generated modes, where two of the modes satisfy a two-pump phase-matching constraint at these cone overlap points, and the rest follow from the single-pump FWM interactions as in \cite{wang_single-step_2017}. Instead, modes form at four spots almost directly above and below each pump. This is the preferred reinforced output because each of the four spatial modes is stimulated by a single-pump FWM relationship, as well as a dual-pump FWM relationship. That is, each mode is firsthand correlated to two others, and secondhand correlated (in two different ways) to the third. See, for example, mode 1 in the inset of Figure \ref{fig:setup}, which is correlated to mode 4 by a single-pump process and mode 3 by a dual-pump process, and secondhand correlated to mode 2 by the path $1\rightarrow4\rightarrow2$ as well as $1 \rightarrow 3 \rightarrow 2$. 
This ``stimulated feedback" between the involved modes in such a manner results in the four-spot configuration being the preferred optical stability in such a pump arrangement.  This suggests that the optimally-squeezed and entangled modes resulting from such a pump configuration are the four modes above and below each pump beam.  These results may be extended to more complicated pump beam setups, where the preferred modes will be those satisfying (the most) direct correlations to one another. 

Like in seeded FWM, the generated photons have a frequency separation of $\sim$3 GHz to the red and blue of the pump beam \cite{zhou_characterizing_2016}, i.e. 6 GHz from each other. The ``probe,'' or lower-frequency photons, are closer to resonance for the D1 line in $^{85}$Rb \cite{steck_rubidium_2008}, and therefore experience non-negligible absorption. 
We find, when injecting a probe-frequency beam generated by a double-pass acousto-optical modulator through the cell at our typical operating temperature of 145$^\circ$C, a transmission of $\tau_p \approx 0.7$, which is a result of both the aforementioned near-resonant absorption as well as the imperfect optical transmission of the glass cell. However, like the FWM cone (but unlike seeded FWM), there is no preferred mode for either frequency and the four modes each possess photons of both ``probe'' and ``conjugate'' frequencies \cite{podhora_nonclassical_2017,zhou_characterizing_2016}. 
Thus, while photons at the different frequencies experience differing degrees of absorption, the ratio of output probe to conjugate photons is uniform across all four modes, and the overall output remains balanced. Note that also, as in any FWM experiment, the vapor temperature will need to be significantly colder in order to achieve squeezing measurements, meaning that this loss due to on-resonant absorption will become negligible.

In the case of undepleted, classical pump beams, we may write the interaction Hamiltonian involving the four spatial modes as:
\begin{align}\hat{H}=i\hbar [\epsilon_a(\hat{a}^\dagger_1 \hat{a}^\dagger_4 -\hat{a}_1 \hat{a}_4)+ \epsilon_b(\hat{a}^\dagger_2 \hat{a}^\dagger_3-\hat{a}_2 \hat{a}_3)+\notag \\
\epsilon_c(\hat{a}^\dagger_1 \hat{a}^\dagger_3 -\hat{a}_1 \hat{a}_3)+
\epsilon_d(\hat{a}^\dagger_2 \hat{a}^\dagger_4 -\hat{a}_2 \hat{a}_4)],\end{align}
where the four modes are labeled $1,2,3,4$ clockwise starting from top left as in Figure \ref{fig:setup}, and $\hat{a}_j$ and $\hat{a}_j^\dagger$s are the usual bosonic annihilation and creation operators. We assume the two pump beams are equal in power and size, so the interaction strengths $\epsilon_i$ ($i=a,b,c,d$) of each single-pump, and each double-pump four-wave mixing process may be taken to be equal to $\epsilon_S$ and $\epsilon_D$, respectively, by symmetry. Then
\begin{align}\hat{H}=i\hbar [\epsilon_S(\hat{a}^\dagger_1 \hat{a}^\dagger_4 -\hat{a}_1 \hat{a}_4+\hat{a}^\dagger_2 \hat{a}^\dagger_3-\hat{a}_2 \hat{a}_3)+\notag \\
\epsilon_D(\hat{a}^\dagger_1 \hat{a}^\dagger_3 -\hat{a}_1 \hat{a}_3+\hat{a}^\dagger_2 \hat{a}^\dagger_4 -\hat{a}_2 \hat{a}_4)].\end{align}

Each photon in this two-pump composition is then directly correlated to two others: one by the single-pump FWM process, and one by the dual-pump FWM process. Note that as long as the pump sizes and powers are equal, each of the four modes is subject to the same interaction, regardless of pump angle, which facilitates a stabilized, balanced output. This is one advantage to this seedless four-mode setup over, for example, the six-mode configuration in \cite{wang_single-step_2017} where equally-bright modes are difficult to achieve. Additionally, these results suggest that the four modes above and below the pump beams are the preferred configuration for detecting optimal degrees of squeezing and entanglement across multiple modes, when two pump beams cross in the present manner.

\section{Conclusion}
In summary, we present a four-mode optical stability arising from a new phase-matched four-wave mixing configuration with multiple pump beams.  Two separate phase-matching constraints are satisfied, and reinforce one another in such a way that the generated photons stimulate each type of phase-matched four-wave mixing process.  These results suggest that the four-mode configuration will exhibit the strongest degree of squeezing and entanglement in experiments with a pair of pump beams crossing at a small angle.  Additionally, we believe that the results could be applied to multi-party entanglement distribution as in \cite{gupta_multi-channel_2016} wherein Gupta et al. generated pairs of correlated random bit streams for secure keys using opposite spatial locations about the single-pump spontaneous FWM ring. 
With the double-pump configuration described here, each photon is correlated to two others rather than one, and no frequency-modulation or probe alignment is necessary. Extending the present results to lower atomic densities (i.e. temperatures) and employing homodyne detection of the four modes should allow for multiple combinations of entangled output states. Lastly, we are hopeful that the present results will be useful in optimizing such entanglement networks generated by similar four-wave mixing processes, by allowing for the prediction of optimal output mode stabilities.

\section*{Acknowledgements} 
We are grateful for discussions with Wenlei Zhang. This material is based upon work supported by the National Science Foundation Graduate Research Fellowship under Grant No. DGE-1154145, as well as the Louisiana State Board of Regents and Northrop Grumman $\emph{NG - NEXT}$.

\bibliography{Zotero3}

~
\newpage

\appendix
\label{sec:add}
\setcounter{figure}{0}
\setcounter{equation}{0}

\makeatletter
\renewcommand{\theequation}{S\@arabic\c@equation}
\renewcommand{\thefigure}{S\@arabic\c@figure}
\makeatletter

\end{document}